\def \SAIT #1 #2 {{\em Mem.\ Soc.\ Astron.\ It.\/} {\bf #1}, #2}
\def \MESS #1 #2 {{\em The Messenger\/} {\bf #1}, #2}
\def \ASTRNACH #1 #2 {{\em Astron. Nach.\/} {\bf #1}, #2}
\def \AAP #1 #2 {{\em Astron. Astrophys.\/} {\bf #1}, #2}
\def \AAL #1 #2 {{\em Astron. Astrophys. Lett.\/} {\bf #1}, L#2}
\def \AAR #1 #2 {{\em Astron. Astrophys. Rev.\/} {\bf #1}, #2}
\def \AAS #1 #2 {{\em Astron. Astrophys. Suppl. Ser.\/} {\bf #1}, #2}
\def \AJ #1 #2 {{\em Astron. J.\/} {\bf #1}, #2}
\def \ANNREV #1 #2 {{\em Ann. Rev. Astron. Astrophys.\/} {\bf #1}, #2}
\def \APJ #1 #2 {{\em Astrophys. J.\/} {\bf #1}, #2}
\def \APJL #1 #2 {{\em Astrophys. J. Lett.\/} {\bf #1}, L#2}
\def \APJS #1 #2 {{\em Astrophys. J. Suppl.\/} {\bf #1}, #2}
\def \APSS #1 #2 {{\em Astrophys. Space Sci.\/} {\bf #1}, #2}
\def \ASR #1 #2 {{\em Adv. Space Res.\/} {\bf #1}, #2}
\def \BAIC #1 #2 {{\em Bull. Astron. Inst. Czechosl.\/} {\bf #1}, #2}
\def \JSQRT #1 #2 {{\em J. Quant. Spectrosc. Radiat. Transfer\/} {\bf #1}, #2}
\def \MN #1 #2 {{\em Mon. Not. R. Astr. Soc.\/} {\bf #1}, #2}
\def \MEM #1 #2 {{\em Mem. R. Astr. Soc.\/} {\bf #1}, #2}
\def \PLR #1 #2 {{\em Phys. Lett. Rev.\/} {\bf #1}, #2}
\def \PASJ #1 #2 {{\em Publ. Astron. Soc. Japan\/} {\bf #1}, #2}
\def \PASP #1 #2 {{\em Publ. Astr. Soc. Pacific\/} {\bf #1}, #2}
\def \NAT #1 #2 {{\em Nature\/} {\bf #1}, #2}

\documentstyle{memsait}
\input epsfig.sty
\begin{opening}
\title{FIRST ASCA RESULTS FOR HIGH-REDSHIFT RADIO-QUIET QUASARS.}
\author{C. VIGNALI$^1$, A. COMASTRI$^2$, M. CAPPI$^3$, 
G.G.C. PALUMBO$^{1,3}$, M. MATSUOKA$^4$}
\institute
{$^1$Dipartimento di Astronomia, Universit\`a di Bologna, Italy\\
$^2$Osservatorio Astronomico di Bologna, Italy\\
$^3$Istituto per le Tecnologie e Studio delle Radiazioni 
Extraterrestri (ITeSRE), Bologna, Italy\\
$^4$The Institute of Physical and Chemical Research (RIKEN), Hirosawa, 
Wako, Japan}
\date{} 
\end{opening}
\def\pn{\par\noindent}

\begin{document}

\oddpagefooter{}{}{} 
\evenpagefooter{}{}{} 
\ 
\bigskip

\begin{abstract}
Here we report the X--ray spectral analysis of 5 high-redshift 
(z$\geq$2) radio-quiet quasars (RQQs) observed with the ASCA satellite. 
A simple power law with Galactic absorption model well fits 
the data. There is no evidence of a reflection component. 
The average spectral slope in the observed 0.5--10 keV energy range, 
$<$$\Gamma$$>$=1.68$\pm{0.09}$ (dispersion $\sigma$$\sim$0.11), 
places the high-z objects of the present sample 
in the lower-$\Gamma$ tail of the RQQs spectral indices distribution. 
\end{abstract}

\section{Introduction}
Early X--ray observations of quasars, both radio-quiet (RQQs) 
and radio-loud (RLQs), have been mainly concerned with low-redshift objects. 
The results achieved with {\it EXOSAT} 
(Comastri et al. 1992) and {\it Ginga} 
(Williams et al. 1992) have indicated that the hard X--ray spectra of RQQs 
are significantly steeper ($\Gamma$$\sim$2) than those of RLQs 
($\Gamma$$\sim$1.6), thus confirming and extending the results obtained 
by {\it Einstein} (Wilkes \& Elvis 1987) in the 0.2--3.5 keV energy range. 
In the last few years, thanks to ASCA capabilities, 
it has been possible to considerably extend the knowledge of the 
hard X--ray spectral properties of high-z RLQs (Cappi et al. 1997), which 
appear as flat ($\Gamma$=1.61$\pm{0.04}$) as the lower-z objects, 
thus giving support to the no-spectral evolution hypothesis. 
On the other hand, the spectral properties of high-z 
RQQs are still widely unknown, given their lower X--ray 
luminosity with respect to RLQs (Zamorani et al. 1981). 
In this context a program has been started to investigate the spectral properties 
of a sample of high-z RQQs (see Table~1), 
choosen among the brightest objects found cross-correlating the V\'eron-V\'eron quasars 
catalogue (V\'eron-Cetty \& V\'eron 1996) with the ROSAT All-Sky Survey 
catalogue of X-ray sources (Voges et al. 1996). Even if not complete and biased towards 
relatively unabsorbed sources, this sample is considered adequate in order to obtain, 
for the first time, a reliable measurement of the X--ray spectral 
properties of high-z RQQs, over the $\sim$ 1.5--30 keV 
energy range (source frame). 

\begin{table}[h]
\centerline{\bf Tab. 1 - The Radio-Quiet Quasars sample}
\begin{center}
\begin{tabular}{|l|c|c|c|c|}
\hline
\multicolumn{1}{l}{\bf Object} &
\multicolumn{1}{c}{z} &
\multicolumn{1}{c}{$N_{{\rm H}_{\rm gal}}^{a}$} &
\multicolumn{1}{c}{$m_{\rm V}$} &
\multicolumn{1}{c}{$R_{\rm L}^{b}$}\\
\hline
{\bf 0040$+$0034} (UM~269) & {\bf 2.00} & 2.45 & 18.0 & $<$ 0.67 \\
{\bf 0300$-$4342} (C~25.36) & {\bf 2.30} & 1.83 & 19.2 & $<$ 1.15 \\
{\bf 1101$-$264} (PG~1101$-$264) & {\bf 2.15} & 5.68 & 16.0 & $<$ -0.22 \\
{\bf 1255$+$3536} (WEE~83) & {\bf 2.04} & 1.22 & 20.6 & $<$ 1.71 \\
{\bf 1352$-$2242} (CTS~327) & {\bf 2.00} & 5.88 & 18.2 & $<$ 0.72 \\
\hline
{\it 1559$+$089} & {\it 2.27} & 3.84 & 16.7 & $<$ 1.00 \\
{\it 1725$+$503} & {\it 2.10} & 2.58 & 20.4 & $<$ 1.63 \\
{\it 1726$+$504} & {\it 1.90} & 2.58 & 19.9 & $<$ 1.43 \\
\hline
\end{tabular}
\end{center}
\vspace{-0.2cm}
\hspace{2.0cm} $^{a}$ In units of 10$^{20}$ cm$^{-2}$, Dickey \& Lockman 1990.
\pn
\hspace{2.0cm} $^{b}$ Radio loudness, defined as 
$R_{\rm L}$ = Log (f$_{\rm 5 GHz}$/f$_{\rm V}$).
\end{table}

\section{ASCA data: reduction and spectral analysis}
The high-z RQQs of the present sample were observed with the ASCA satellite (Tanaka, Inoue 
\& Holt 1994) during the AO4 and AO5 phases for about 40 ks each, except for 1101$-$264 
(20 ks). 
In the following, we'll refer to SIS (solid-state spectrometers) 
data only given their better statistics compared to GIS (gas scintillation 
spectrometers) one.\\
A simple power law plus photoelectric absorption 
was fitted to the data, with the column density left initially free and then fixed to 
the Galactic value (Table~2). 
No excess absorption is required by the data. Then 
spectral fittings with two separate absorbers, one located at z=0 
and fixed to the Galactic value and one at the quasar redshift, 
were performed. Again there is no need of excess absorption. 

\begin{table}[h]
\centerline{\bf Tab. 2 - Spectral results with a single power law model}
\begin{center}
\begin{tabular}{|l|c|c|c|c|c|}
\hline
\multicolumn{1}{l}{\bf Object} &
\multicolumn{1}{c}{$N_{\rm H}$} &
\multicolumn{1}{c}{$\Gamma$} &
\multicolumn{1}{c}{$\chi^{2}$/dof} &
\multicolumn{1}{c}{$F_{2-10 keV}$} &
\multicolumn{1}{c}{$L_{2-10 keV}$} \\
\multicolumn{1}{l}{} &
\multicolumn{1}{c}{(10$^{20}$ cm$^{-2}$)} &
\multicolumn{1}{c}{} &
\multicolumn{1}{c}{} &
\multicolumn{1}{c}{(10$^{-13}$ cgs)} &
\multicolumn{1}{c}{(10$^{46}$ cgs)} \\
\hline
{\bf 0040$+$0034} & $\equiv$$N_{{\rm H}_{\rm gal}}$ & 1.63$\pm{0.07}$ & 109/121 & 14 & 2.4 \\
 & 9.74$^{+5.46}_{-5.15}$ & 1.79$^{+0.15}_{-0.14}$ & 104/120 & & \\
\hline
{\bf 0300$-$4342} & $\equiv$$N_{{\rm H}_{\rm gal}}$ & 1.75$\pm{0.14}$ & 81.9/79 & 4.0 & 1.1 \\
 & $<$ 6.84 & 1.69$^{+0.24}_{-0.13}$ & 81.5/78 & & \\
\hline
{\bf 1101$-$264} & $\equiv$$N_{{\rm H}_{\rm gal}}$ & 1.83$^{+0.24}_{-0.22}$ 
& 28.2/27 & 2.7 & 0.7 \\
 & 19.5$^{+19.5}_{-15.5}$ & 2.19$^{+0.58}_{-0.47}$ & 26/26 & & \\
\hline
{\bf 1255$+$3536} & $\equiv$$N_{{\rm H}_{\rm gal}}$ & 1.56$^{+0.06}_{-0.07}$ & 131/129 & 
12 & 2.1 \\
& $<$ 6.62 & 1.59$^{+0.12}_{-0.11}$ & 131/128 & & \\
\hline
{\bf 1352$-$2242} & $\equiv$$N_{{\rm H}_{\rm gal}}$ & 1.65$^{+0.13}_{-0.14}$ & 
58.9/70 & 6.0 & 1.0 \\
 & $<$ 17.9 & 1.66$^{+0.25}_{-0.23}$ & 58.9/69 & & \\
\hline
{\it 1559$+$089} & $\equiv$$N_{{\rm H}_{\rm gal}}$ & 1.8 & & $<$ 0.6 & $<$ 0.16 \\
{\it 1725$+$503} & $\equiv$$N_{{\rm H}_{\rm gal}}$ & 1.8 & & $<$ 1.2 & $<$ 0.28 \\
{\it 1726$+$504} & $\equiv$$N_{{\rm H}_{\rm gal}}$ & 1.8 & & $<$ 1 & $<$ 0.19 \\
\hline
\end{tabular}
\end{center}
\pn
A Friedmann cosmology with $H_{0}$=50 Km s$^{-1}$ Mpc$^{-1}$ and $q_{0}$=0.5 is assumed.
\pn
Intervals are at 90\% confidence for one interesting parameter 
($\Delta\chi^{2}$=2.71).
\pn
For the last three quasars of the list the data have been obtained 
as 3 $\sigma$ upper limits. 
\end{table} 

\pn 
Although the sources are rather faint in ASCA, 
the derived parameters weakly depend on the assumed 
background normalizations. 
A simple unweighted 0.5--10 keV (observer frame) spectral slopes mean gives 
$<$$\Gamma$$>$=1.68$\pm{0.09}$ ($\sigma$$\sim$0.11) and 
$<$$\Gamma$$>$=1.78$\pm{0.17}$ ($\sigma$$\sim$0.24) with 
$N_{\rm H}$=$N_{{\rm H}_{\rm gal}}$ and $N_{\rm H}$ free, respectively. 
Moreover, the lack of any intrinsic absorption makes 
high-z RQQs quite different with respect to high-z RLQs, since for the latter class clear 
indications of excess absorption have been found (Elvis et al. 1994, Cappi et al. 1997). 
There is no evidence of spectral flattening attributed to a reflection component and 
only for 1101$-$264 a Fe\,K$\alpha$ emission line has been marginally detected 
(E=6.51$\pm{0.25}$ keV, EW$\simeq$690$\pm{560}$ eV in the source frame). 
Then, a co-added spectrum 
has been computed rescaling all the objects at z=2. 
The best-fit model (Fig.~1) has ben achieved with 
a power law ($\Gamma$=1.71$^{+0.08}_{-0.07}$) plus absorption again consistent with 
the average Galactic value (Fig.~2). Thermal models can be ruled out on the basis of 
the high temperatures obtained (physically difficult to be interpreted) and worse $\chi^{2}$. 
The characteristic features of reprocessing gas, such 
as the high energy ``hump" and the iron emission line, 
have not been revealed. Moreover, tight constraints have been obtained on both 
the Fe K$\alpha$ EW ($<$133 eV at z=2) and the reflection component (R$<$1.2, where R is 
the relative amount of reflection compared to the primary power law). 
Since this component peaks at 20--30 keV in the quasars frame, 
the present ASCA observations should reveal it, if really present. 
%
\begin{figure}
\parbox{6truecm}
{\epsfig{figure=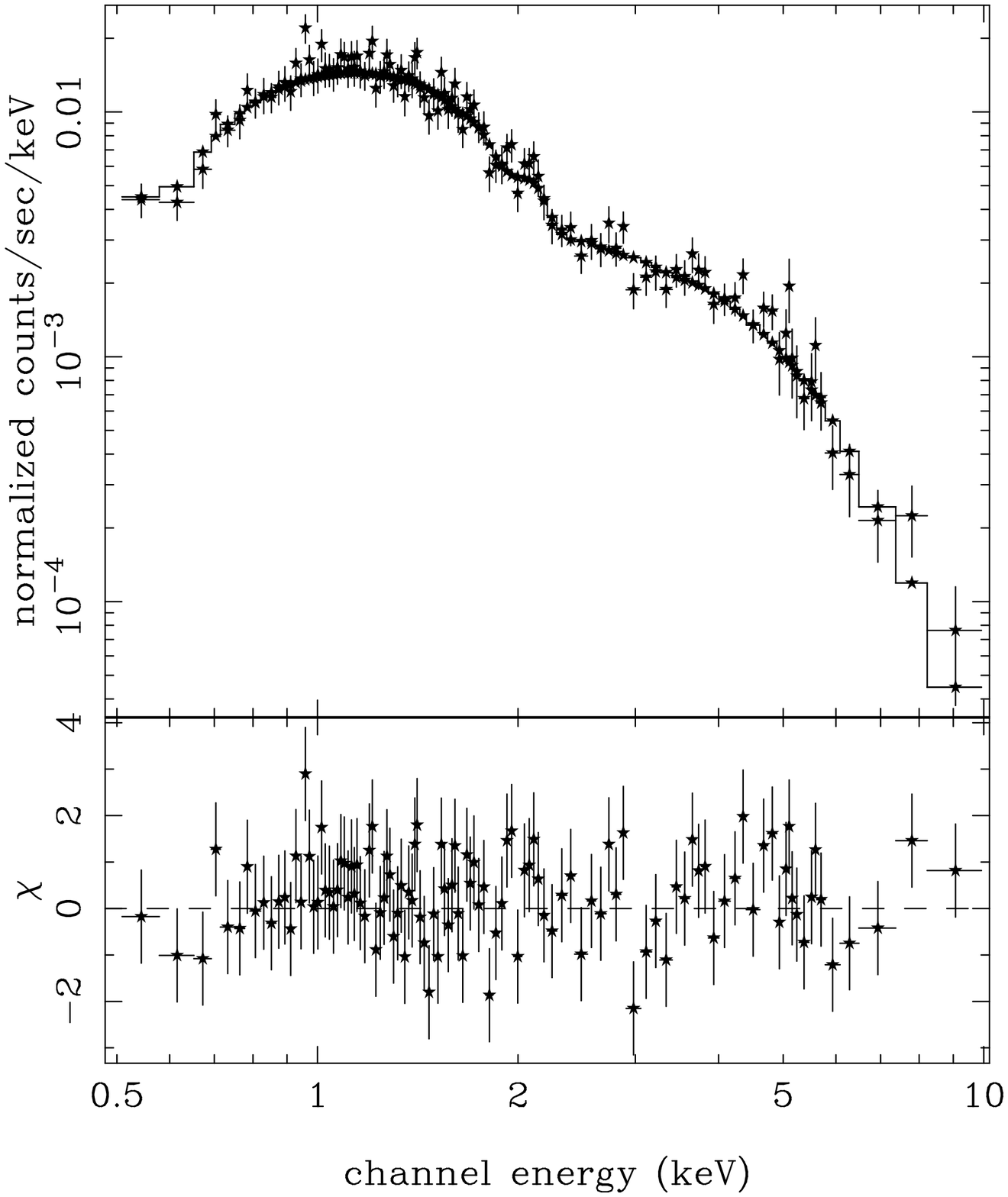,height=6.3cm,width=6.3cm,angle=0}
{\vskip -0.3cm \footnotesize Fig.~1: Plot of the co-added spectrum obtained re-scaling the 
energy scale of each quasar to z=2.}}
\  \hspace{0.8truecm}     \
\parbox{6truecm}
{\vskip 0.35cm \epsfig{figure=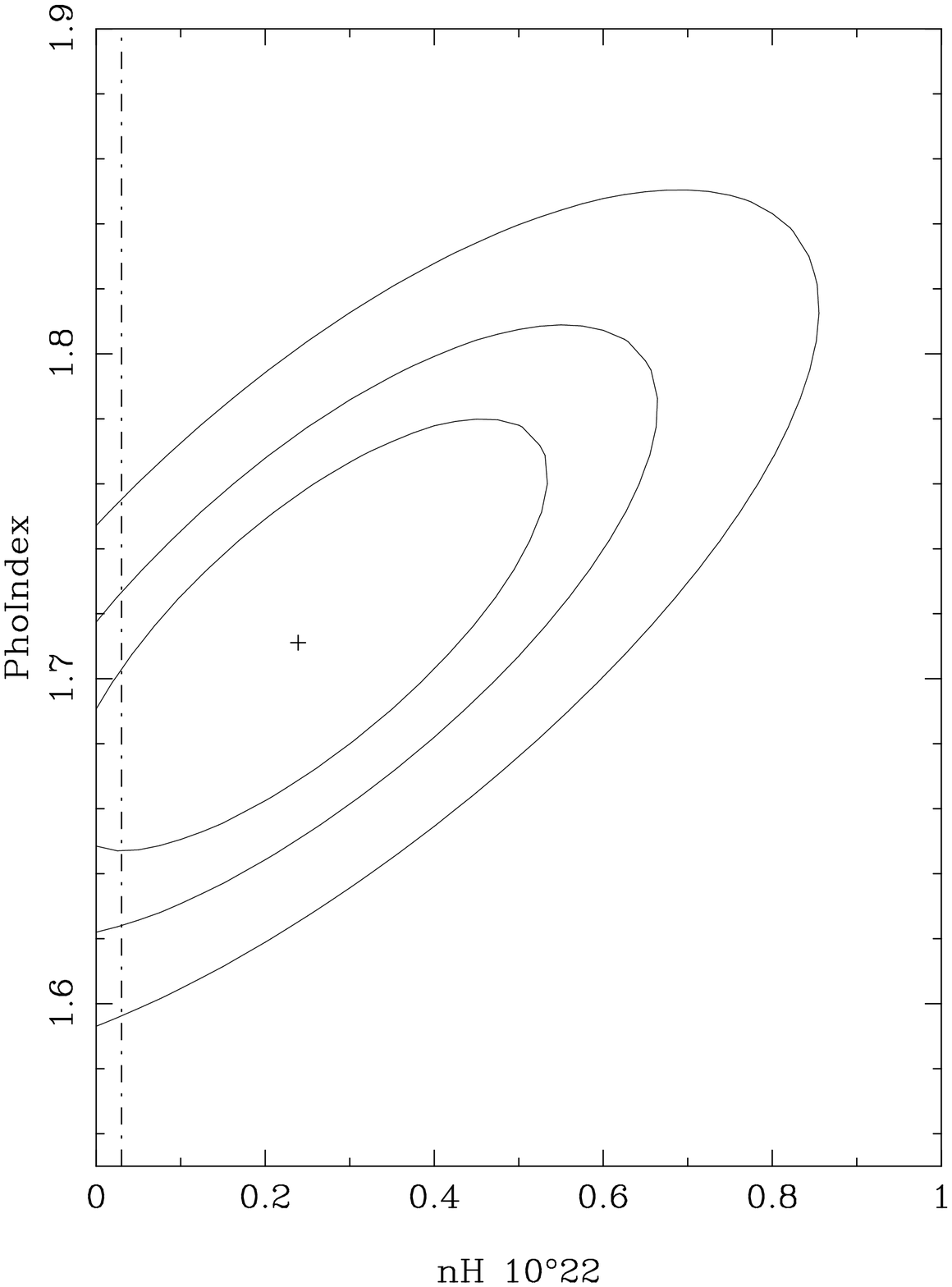,height=5.6cm,width=5.6cm,angle=0}
{\footnotesize Fig.~2: 68, 90 and 99\% $N_{\rm H}$ -- $\Gamma$ 
confidence contours. The dash-dotted line marks the Galactic absorption value 
($N_{\rm H}$$\sim$3$\times$10$^{20}$ cm$^{-2}$).}}
\end{figure}
\section{Discussion}
The most important result achieved from the present analysis is that all 
the quasars have similar slopes. 
The average spectral slope is $<$$\Gamma$$>$=1.68$\pm{0.09}$ with a dispersion $\sigma$$\sim$0.11. 
The flatter X--ray spectra for high-z RQQs with respect to low-z objects, imply that 
{\bf either} we are sampling the RQQs lower tail of the photon indices distribution (Fig.~3) 
{\bf or} the primary emission mechanism is really different at high redshift. 
The imprints of cold matter, 
both in transmission (absorption) and in reflection (Fe K$\alpha$ emission line 
and Compton ``hump"), have not been found. 
This result may be related to the quasars accretion rate, 
i.e. they can be accreting close to their Eddington limit. 
If this is the case, in high-luminosity objects ($L_{\rm X}$$\approx$$10^{46}$ erg s$^{-1}$), 
the disc may highly ionized and the lack of the iron line may be caused either 
by complete ionization of the iron atoms (\.{Z}ycki \& Czerny 1994) or by 
resonant trapping of the line photons. 
Moreover, the high degree of ionization is likely to smooth out also the reflection component. 
In fact, as suggested by Nandra et al. (1997) for 
high-ionization environments, the Compton reflection could be seen without 
suffering absorption in the disc. This reduces the constrast with the underlying continuum, 
resulting in an apparently weak Compton ``hump". 
Alternatively, there may be different environmental and/or geometrical properties for
RQQs with respect to Seyfert galaxies.
\begin{figure}
\epsfysize=5cm 
\hspace{3.5cm}\epsfbox{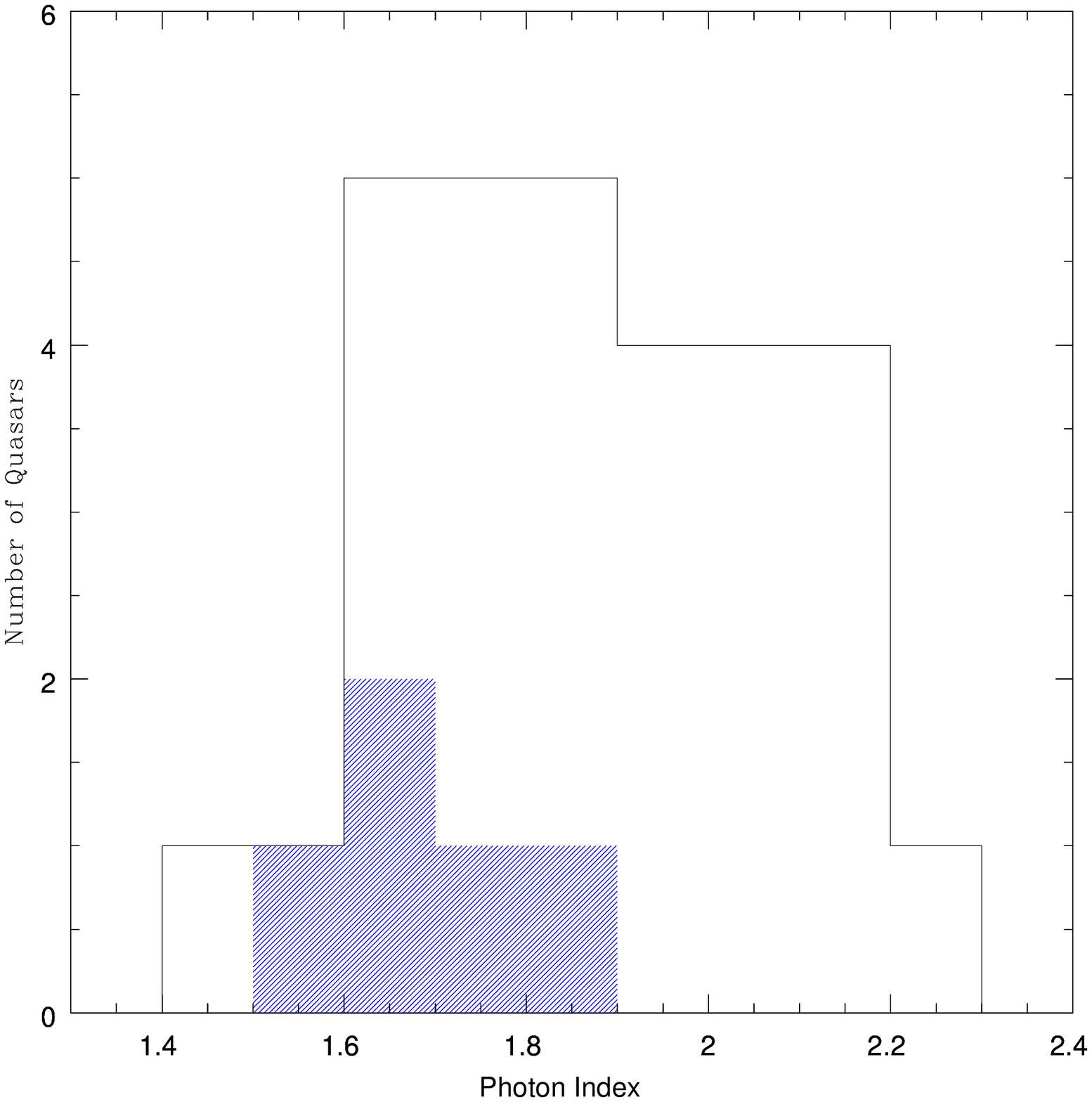}
\pn
{\footnotesize Fig.~3: 
A histogram distribution of X--ray spectral slopes for the our RQQs (shaded area) 
compared to literature samples of RQQs.}
\end{figure}
%
%



\end{document}